\documentclass[amsmath,amssymb,aps,prl,twocolumn,showkeys,superscriptaddress,10pt,floatfix,showpacs]{revtex4-1}

\usepackage{graphicx}
\usepackage{amssymb}
\usepackage{dcolumn}
\usepackage{bm}
\usepackage{hyperref}
\usepackage{color}
\usepackage{siunitx}

\usepackage{xcolor}
\usepackage{float}

\hypersetup{
    colorlinks,
    linkcolor={red!50!black},
    citecolor={blue!50!black},
    urlcolor={blue!80!black}
}

\begin{document}

\title{Possible pressure-induced topological quantum phase transition in the nodal line semimetal ZrSiS}

\author{D. VanGennep}
\author{T. A. Paul}
\author{C. W. Yerger}
\affiliation{Department of Physics, University of Florida, Gainesville, FL 32611}
\author{S.\ T.\ Weir}
\affiliation{Physics Division, Lawrence Livermore National Laboratory, Livermore, CA 94550, USA}
\author{Y.\ K.\ Vohra}
\affiliation{Department of Physics, University of Alabama at Birmingham, Birmingham, AL, 35294}
\author{J. J. Hamlin}
\email{jhamlin@ufl.edu}
\affiliation{Department of Physics, University of Florida, Gainesville, FL 32611}


\begin{abstract}
ZrSiS has recently gained attention due to its unusual electronic properties: nearly perfect electron-hole compensation, large, anisotropic magneto-resistance, multiple Dirac nodes near the Fermi level, and an extremely large range of linear dispersion of up to $\sim$ 2 eV.
We have carried out a series of high pressure electrical resistivity measurements on single crystals of ZrSiS.
Shubnikov-de Haas measurements show two distinct oscillation frequencies.
For the smaller orbit, we observe a change in the phase of $\sim$0.5, which occurs between $0.16-0.5$ GPa.
This change in phase is accompanied by an abrupt decrease of the cross-sectional area of this Fermi surface.
We attribute this change in phase to a possible topological quantum phase transition.
The phase of the larger orbit exhibits a Berry phase of $\pi$ and remains roughly constant up to $\sim$2.3 GPa.
Resistivity measurements to higher pressures show no evidence for pressure-induced superconductivity to at least $\sim \SI{20}{GPa}$.
\end{abstract}

\maketitle

\section{Introduction}
\label{sec:Introduction}
Three-dimensional (3D) Dirac semimetals are a new class of quantum materials that host two sets of linear, doubly degenerate bands which cross at so-called Dirac points.
Breaking time reversal or inversion symmetry lifts the degeneracy of these bands, resulting in singly degenerate band crossings referred to as Weyl nodes~\cite{Yan2017_1}.
Few materials have been experimentally verified as hosts of these exotic band dispersions, which are analogous to 3D versions of graphene.
Both Cd$_3$As$_2$ and Na$_3$Bi have shown evidence of 3D symmetry-protected Dirac cones in the bulk~\cite{Liu2014_1,Liu2014_2}.
In both materials, the energy range in which the band dispersion remains linear is quite small, on the order of 0.01-0.1 eV, which can make them challenging to systematically study because impurities, defects, and pressure can alter the chemical potential away from the linear regime~\cite{Schoop_2016}.
ZrSiS hosts multiple Dirac cones and has been shown to exhibit an unusually robust linear dispersion (up to $\sim$ 2 eV in part of the Brillouin zone)~\cite{Singha_2017}.
ZrSiS is also chemically stable (as opposed to Na$_3$Bi) and non-toxic (as opposed to Cd$_3$As$_2$).
Together, these features make ZrSiS a promising system for studying the physics of 3D Dirac/Weyl fermions. 

The crystal structure of ZrSiS can be described as layered, containing quintuple layers of S-Zr-Si-Zr-S, with a PbFCl structure and space group P4/nmm (No.\ 129), lattice parameters of $a = b = 3.5440$~\AA, $c = 8.0550$~\AA, and volume $V = 101.17$~\AA$^3$~\cite{Sankar_2017,Lv_2016}.
Polycrystalline samples were first synthesized via solid state reaction by Haneveld \textit{et al.}~\cite{Haneveld_1964}.
Millimeter-sized, high quality single crystals can be grown via vapor transport of polycrystalline source material, using iodine as a vapor transport agent~\cite{Sankar_2017,Lv_2016}. 

Both angle-resolved photoemission spectroscopy (ARPES) experiments and electronic structure calculations have shown evidence of a Dirac line node phase, a diamond-shaped Fermi surface at the Brillouin zone center $\mathbf{(\Gamma)}$ point, an ellipsoidal-shaped Fermi surface at the \textbf{M} point, and small electron-like pockets at the \textbf{X} point~\cite{Sankar_2017,Schoop_2016,Wang_2016_1,Neupane_2016,Lv_2016,Zhou_2017}.  
The ambient pressure Fermi surface has also been well characterized via quantum oscillation measurements~\cite{Sankar_2017,Wang_2016_1,Matusiak_2017,Hu_2017_2,Zhang_2017,Singha_2017,Ali_2016,Lv_2016}.
Most reports provide evidence for two distinct oscillation frequencies when the magnetic field is aligned along the crystallographic $c$-axis.
Ali \textit{et al.}~\cite{Ali_2016} reported that the phase of the high frequency oscillation goes through a sharp transition as a function of the angle of the applied magnetic field.
Matusiak \textit{et al.}~\cite{Matusiak_2017} found the thermoelectric response in ZrSiS to be a more sensitive Fermi surface probe than Shubnikov-de Haas (SdH) or de Haas-van Alphen (dHvA) quantum oscillations, observing a total of five distinct oscillation frequencies, with some oscillations still resolvable at 100 K. 

One of the unique features of ZrSiS that has gained much attention is the large, anisotropic magneto-resistance, which can be as high as $1.4\times 10^5~\%$ at 2 K and 9 T.
The magneto-resistance is maximized when the magnetic field is aligned along the [011] axis~\cite{Singha_2017}.
Hall measurements suggest ZrSiS exhibits a nearly perfect electron-hole compensation ratio of $\sim$ 0.94~\cite{Zhang_2017,Lv_2016}.
Lv \textit{et al.}~\cite{Lv_2016} suggest the unusual magneto-resistive properties of ZrSiS can be attributed to the electron-hole compensation as well as the open orbital Fermi surface.
The relativistic phenomenon of massless Dirac fermions known as the Adler-Bell-Jackiw~\cite{Adler1969,Bell1969} chiral anomaly has also been observed in ZrSiS~\cite{Singha_2017}.

Recently, Singha \textit{et al.}~\cite{Singha2018_1} studied the effect of pressure on the lattice dynamics and electron-phonon coupling in ZrSiS, which revealed two pressure-induced structural phase transitions near 3.7 and 18.7 GPa.
ZrSiS was also found to exhibit tip-induced superconductivity coexisting with the preserved topological properties in a point-contact electrical study~\cite{Aggarwal2018_1}.
The authors suggest the tip-induced superconductivity arises due to an increase in the density of states near the Fermi level due to the presence of the Ag point contact.
The same work also reported an absence of pressure-induced superconductivity in ZrSiS to at least \SI{8}{GPa}. The base temperature for the high-pressure measurements was not stated.

How the unique electronic properties of ZrSiS might evolve under pressure is, at present, largely an open question.
In this work, we report measurements of Shubnikov-de Haas (SdH) oscillations in single crystals of ZrSiS to hydrostatic pressures of $\sim 2.5$~GPa.
We also report the results of electrical resistivity measurements to $\sim 27$~GPa.

\section{Experimental methods}
\label{sec:Methods}
Single crystals of ZrSiS were grown by solid state reaction followed by chemical vapor transport, using the prescription detailed in Ref.~\cite{Lv_2016}. 
The crystal structure was characterized via powder x-ray diffraction, and Rietveld refinement of the data gave lattice parameters of $a = b = 3.55$~\AA, and $c = 8.06$~\AA, which are consistent with literature values~\cite{Sankar_2017,Lv_2016}.
Rocking curve measurements of un-cleaved single crystals present a single sharp peak indicating high crystal quality.
Small pieces of sample with dimensions of about $500\,\mathrm{\mu m} \times 500\,\mathrm{\mu m} \times 100\,\mathrm{\mu m}$ were cut from a larger crystal.
For the low-pressure SdH measurements, Pt wires were connected to the samples using EPO-TEK H20E conductive epoxy.
The samples were then mounted to the wire and fiber optic feed-throughs of a Teflon-capsule piston-cylinder type pressure cell constructed of MP35N alloy.
The pressure was calibrated at both room temperature and the lowest temperature reached using the fluorescence of the R1 peak of a small ruby chip~\cite{Chijioke2005}.
Daphne 7474 oil was used as the pressure-transmitting medium surrounding the sample~\cite{murata_2008_1}.
At room temperature, Daphne 7474 does not solidify until 3.7~GPa, which is beyond the range of the Teflon-capsule cell measurements.
Four-wire resistance measurements were performed in the crystalline $ab$-plane using either a Quantum Design PPMS resistance bridge or a Lakeshore 370 resistance bridge.
Magnetic fields were applied along the $c$-axis.
Samples 2, 3, and 4 were studied at ambient pressure only, while samples 1 and 8 were subjected to SdH measurements under pressure. 

The higher-pressure resistance measurements were carried out on single crystals of ZrSiS (samples 5 and 6) in a gas membrane-driven diamond anvil cell.
The pressure was measured using the fluorescence of the $R_1$ peak of small ruby spheres placed next to the sample~\cite{Chijioke2005}.
One of the diamonds used was a designer diamond anvil~\cite{weir_2000_1}.
Resistance was measured in the crystalline $ab$-plane by a Lakeshore Model 370 AC resistance bridge using the four-probe van der Pauw method with currents of $\leq 1\,\mathrm{mA}$.
Quasihydrostic, soft, solid steatite was used as the pressure-transmitting medium.
Additional details of the high pressure methods are available in Ref.~\cite{VanGennep2017}.

\section{Results}
\label{sec:Results}
\begin{figure}
    \includegraphics[width=\columnwidth]{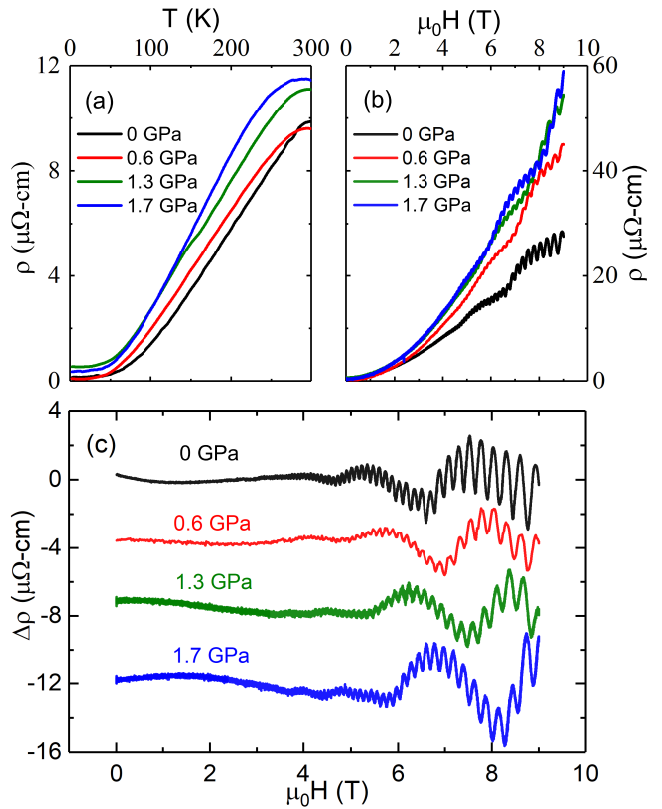}
    \caption{(a) Resistivity vs temperature, (b) resistivity vs magnetic field measured at 2 K, and (c) oscillatory part of the resistivity vs magnetic field at various pressures for sample 8 measured at 2 K. The data have been vertically offset for clarity. The magnetic field was applied parallel to the crystallographic \textit{c}-axis.
Fermi surface parameters derived from analyzing this and other data are presented in Fig.~\ref{fig:fig2}.}
 \label{fig:fig1}
\end{figure}

Hydrostatic pressure measurements of the electrical resistivity of ZrSiS are summarized in Fig.~\ref{fig:fig1}.
In all samples, we find that pressure tends to make the room-temperature resistivity increase.
Pressure produces an increase the magnitude of the high-field resistivity (see Fig.~\ref{fig:fig1}b).
The magnetic field in all of these measurements was applied parallel to the crystallographic $c$-axis, while the resistivity was measured in the crystalline $ab$-plane.

The two SdH frequencies observed in all samples correspond well to the frequencies previously reported at ambient pressure \cite{Singha_2017,Sankar_2017,Wang_2016_1,Matusiak_2017,Hu_2017_2,Zhang_2017,Ali_2016}.
The data agree well with both the oscillation frequency and the phase of each oscillation obtained from the LL fan diagram at ambient pressure.
Thermoelectric measurements indicate several oscillation frequencies which we did not observe in our measurements and were also not observed in other dHvA and SdH experiments~\cite{Matusiak_2017}. 

\begin{figure*}
    \includegraphics[width=0.95\columnwidth]{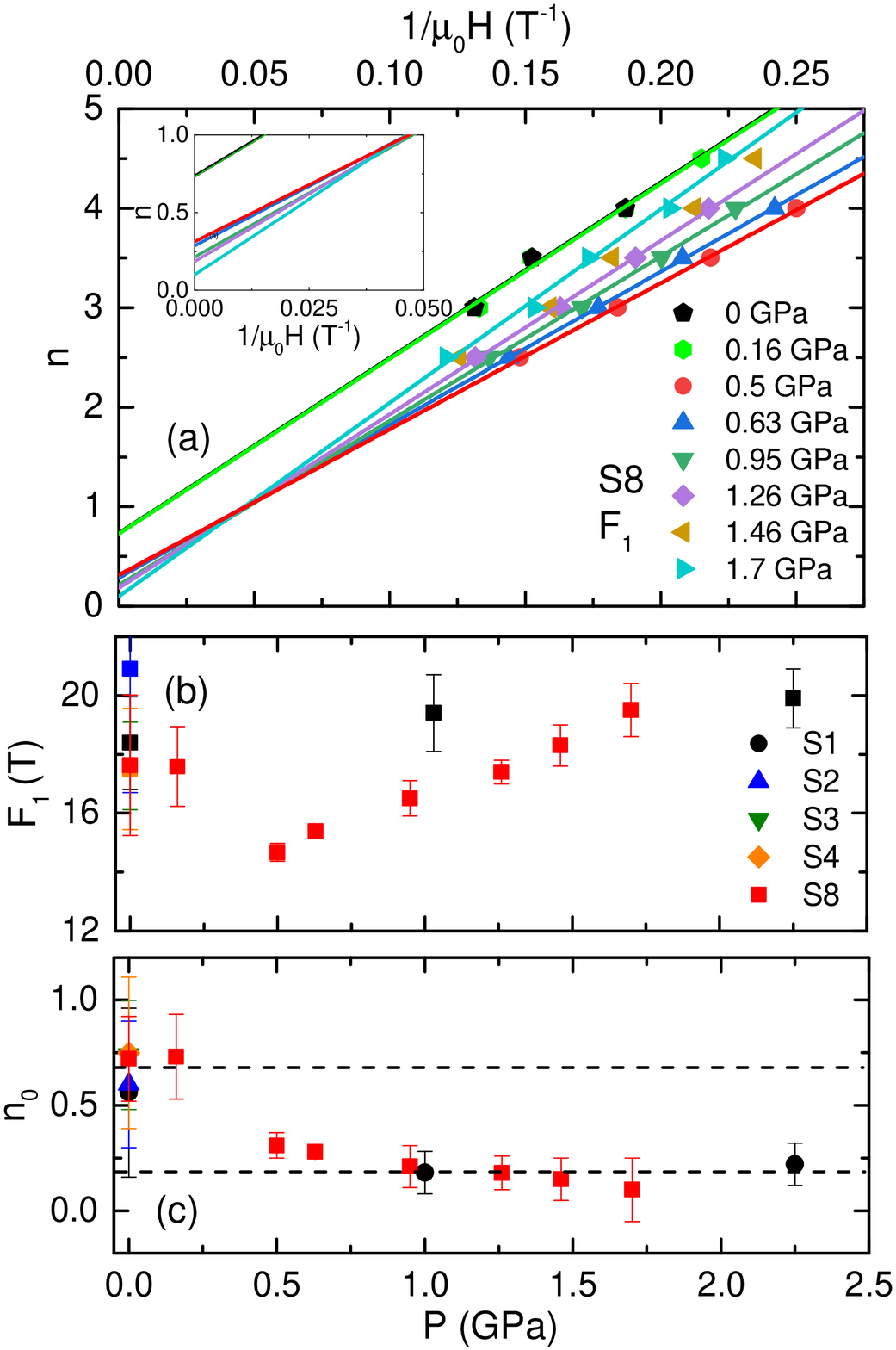}
    \hspace{\columnsep}
    \includegraphics[width=0.95\columnwidth]{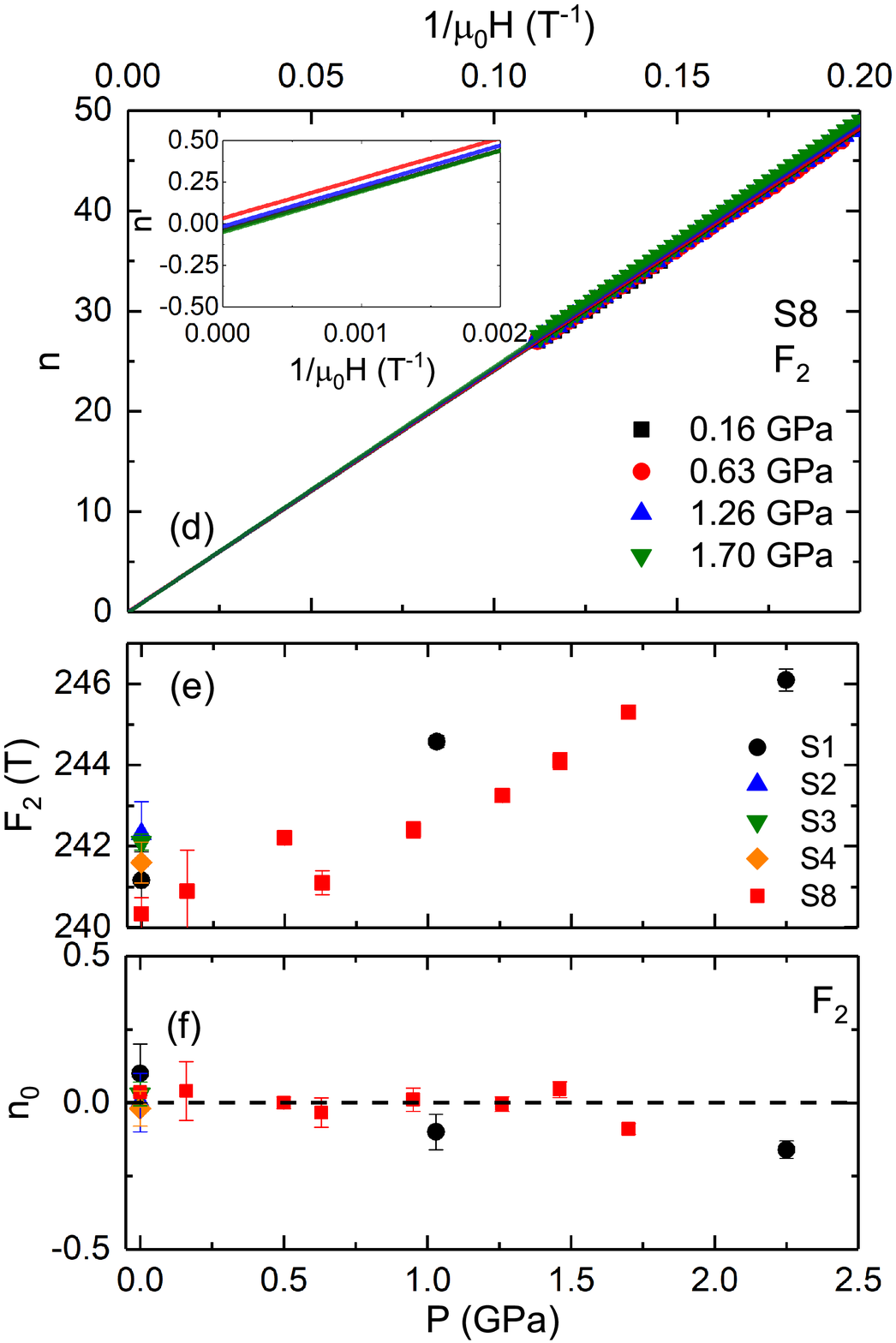}
    \caption{(left) Several Fermi surface parameters of ZrSiS at 2 K as a function of pressure for the smaller Fermi surface, with an oscillation frequency $F_1$. (a) Landau level fan diagram with selected pressures from sample 8. (b) Oscillation frequency $F_1$ for all samples as a function of pressure, (c) n-intercept, $n_0$, of the LL fan diagram of $F_1$. This Fermi surface is known to be 3D and fairly isotropic. The intercept $n_0$ discontinuously drops by $\sim$0.47 between 0.16-0.5 GPa, which could suggest that the Berry phase of this orbit changes by a factor of $\pi$. The dotted lines indicate the average values of $n_0$ below and above the transition pressure, which are 0.66 and 0.19, respectively. The nature of this change in phase is discussed in more detail in the text. (right) Several Fermi surface parameters of ZrSiS at 2 K as a function of pressure for the larger Fermi surface, with frequency $F_2$. (d) Landau level fan diagram with selected pressures from sample 8. (e) Oscillation frequency $F_2$ for all samples as a function of pressure, (f) n-intercept of the LL fan diagram of the larger orbit, which exhibits a Berry phase of $\pi$ and stays roughly constant.}
 \label{fig:fig2}
\end{figure*}

Landau quantization of electronic states gives rise to SdH quantum oscillations, which can be described by the Lifshitz-Kosevich (LK) relation.
The oscillatory part of the LK expression is given by:
\begin{equation}
\Delta\sigma_{xx} \propto cos[2\pi(F/B+\phi)],
\end{equation}
where $B$ is the magntiude of the magnetic field, $F$ is the frequency of the oscillation, and $\phi$ is the phase shift, which encodes information about the topography of the Fermi surface~\cite{Shoenberg1984}.
To identify the phase shift, the Landau indices where $F/B+\phi$ takes on integral values, $n$, need to be identified from the magneto-resistance.
A plot of $n$ vs $1/B$, referred to as a Landau level (LL) fan diagram, then extrapolates to the phase shift on the $n$-axis, which we call $n_0$.
Reference~\cite{Wang_2016_3} provides a useful description of the QO phases observed in 3D topological semimetals. 

Analysis of the quantum oscillation data follows prescriptions described in Refs.~\cite{Shoenberg1984,Ando_2013_1}.
The results of this analysis are summarized in Fig.~\ref{fig:fig2}.
Two distinct frequencies of SdH quantum oscillations are visible in the raw data.
While $\rho_{xx}$ is roughly parabolic as a function of field, subtraction of the non-oscillatory part of the magneto-resistance was performed by fitting the raw data to a 5th-order Chebyshev polynomial.
If only the data above \SI{1}{T} are used, a simple quadradic background subtraction results in the same conclusions regarding LL assignments.
We refer to the small oscillation frequency ($\sim$ 16 T) as $F_1$ and the large frequency ($\sim$ 240 T) as $F_2$.
The cyclotron masses associated with the oscillations are indicated by $m_1^*$ and $m_2^*$, respectively.

In order to determine the correct phases from SdH quantum oscillations, one should make sure that integral values of $n$ are being assigned to minima in $\Delta \sigma_{xx}$. 
Since $\sigma_{xx}= \rho_{xx}/(\rho_{xx}^2+\rho_{xy}^2)$, this could correspond to either maxima or minima in $\Delta\rho_{xx}$, depending on the ratio: $\left|\rho_{xx}/\rho_{xy}\right|$. 
If this ratio is much larger (smaller) than one, integral values of $n$ should occur when $\Delta\rho_{xx}$ is a maximum (minimum)~\cite{Ando_2013_1}.
We performed Hall measurements at ambient pressure and found that  $\left|\rho_{xx}/\rho_{xy}\right| = 17.5$ at 2 K, 9 T for S8, indicating that integral values of $n$ should be assigned to maxima in $\Delta\rho_{xx}$.
Hall measurements were not performed under pressure, but we observe that $\rho_{xx}$ increased by a factor of $\sim$2 from 0-1.7 GPa, measured at 2 K, 9 T for S8. 
This means that $\left|\rho_{xy}\right|$ would have to increase by a factor of $\sim$35 over this pressure range in order to modify our assignment of LLs.
This is highly improbable given that the oscillation frequencies only change by $\sim$15 \% and $\sim$2 \% over the entire pressure range for the small and large Fermi surfaces, respectively, indicating small changes in the carrier densities for these Fermi surfaces.

Figure~\ref{fig:fig2} contains a summary of the SdH oscillation data.
Fig.~\ref{fig:fig2}a,d show the LL fan diagrams for each set of oscillations at various pressures.
Linear fits to the fan diagrams allow us to extrapolate $F$ from the slope (Fig.~\ref{fig:fig2}b,e) and $n_0$ from the $n$-intercept (Fig.~\ref{fig:fig2}c,f).
From ambient pressure to \SI{2.3}{GPa}, $F_2$ increases by $\sim$ 2\%.
The lower frequency, $F_1$ appears to show an abrupt drop below \SI{0.5}{GPa} and then increases at higher pressures.
We considered the possibility that this drop in frequency was due to a small tilt in the sample from the application of pressure, but this could not be the case due to the highly 3D nature of this orbit~\cite{Zhang_2017}.
Temperature dependent data were collected under pressure for sample 1 taken at 7 different temperatures ranging from 2 - 30 K. 
These data revealed a small monotonic decrease in $m_2^*$ from values of 0.17, 0.14, and 0.13 $m_0$ at 0, 1.0, and 2.3 GPa, respectively, which agree well with the literature values at ambient pressure~\cite{Matusiak_2017}.
It was not possible to reliably determine the cyclotron masses of the smaller orbit due to a small number of low frequency oscillations being resolvable for S1. Reference values for $m_1^*$ range between 0.1 - 0.14 m$_0$ at ambient pressure~\cite{Matusiak_2017}.
The data show that $n_0$ for the phase of the large orbit remains constant up to $\sim$ 2.3 GPa, while $n_0$ for the small orbit seems to exhibit an abrupt change between 0.16 - 0.5 GPa.
This nature of this change in phase and its possible significance is discussed in further detail below.

\begin{figure}
	\centering
    \includegraphics[width=0.9\columnwidth]{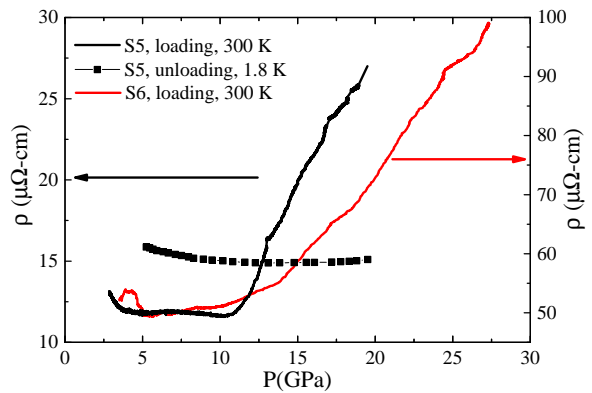}
    \caption{Electrical resistivity of ZrSiS in the $ab$-plane as a function of applied pressure for samples 5 and 6. Clear changes in the slope of $\rho$ vs P occur between $11-14.5$ GPa. Pressure was released at $\sim$ 1.8 K for sample 5.}
 \label{fig:fig3}
\end{figure}

Figure~\ref{fig:fig3} shows the results of electrical resistivity measurements to pressures as high as \SI{27}{GPa}.
For both samples 5 and 6, data was first collected during pressure application at room temperature, where a slope change near $12-\SI{15}{GPa}$ is apparent.
This corresponds roughly to the same pressure at which previous measurements showed changes in the Raman spectrum but is somewhat lower than the pressure where a monoclinic phase first appears (\SI{19}{GPa})~\cite{Singha2018_1}.
For sample 5, after reaching \SI{20}{GPa} at room temperature, the the cell was cooled to \SI{1.8}{K}.
At \SI{20}{GPa}, the sample showed metallic behavior ($d\rho/dT > 0$).
Pressure was then released at low temperature.
During low temperature unloading, the resistance remained roughly constant, which may be related to some hysteresis in the structural transition.
The low temperature unloading data indicate ZrSiS is not superconducting down to \SI{1.8}{K} at these pressure, where ZrSiS has been reported to adopt orthorhombic and monoclinic crystal structures~\cite{Singha2018_1}.

\section{Discussion}
\label{sec:Discussion}
At ambient pressure, it is generally agreed upon that the phase of the low frequency oscillation ($\sim5/8$) corresponds to a topologically nontrivial orbit~\cite{Singha_2017,Hu_2017_2,Zhang_2017,Wang_2016_1}, although band structure calculations tend to lack evidence of this orbit~\cite{Wang_2016_1,Li2018}. For 3D Dirac/Weyl semimetals, phases of $\pm$5/8 and $\pm$1/8 can both be observed from topologically nontrivial orbits~\cite{Wang_2016_3}. Thus, there is a question of whether the observed pressure-driven change in phase in this work corresponds to a topologically nontrivial-nontrivial transition, or a nontrivial-trivial transition. A concrete answer to this question is difficult to determine without a better theoretical understanding of this orbit. Below, we argue that this transition is likely a nontrivial-trivial transition.

The pressure-induced change in the phase of the low frequency oscillation exhibited by ZrSiS (see Fig.~\ref{fig:fig2}c) closely resembles the behavior of Cd$_3$As$_2$ under pressure~\cite{Zhang2017_2}.
At $\sim$1.3 GPa, Cd$_3$As$_2$ shows a sudden change in the phase factor for one of the oscillations accompanied by an abrupt shrinkage of the Fermi surface cross-sectional area. These features are very similar to what we observe in ZrSiS between $\sim$0.16-0.5 GPa. In the case of Cd$_3$As$_2$, the change in phase factor was attributed to pressure-driven node-pair annihilation which results from shifting the Dirac nodes toward the center of the Brillouin zone and eventually introducing a nonzero gap in the energy spectrum~\cite{Zhang2017_2}.
This picture was supported by first principles calculations.
Crucially, x-ray diffraction measurements showed the changes in phase-factor are not due to a change in crystal structure, but are instead purely electronic in nature.
X-ray diffraction measurements under pressure~\cite{Singha2018_1} demonstrate that ZrSiS remains in the ambient pressure crystal structure to pressures above \SI{2.5}{GPa}.
This result confirms that the apparent change in phase observed in the present work is not a consequence of a structural transformation.

Other than the annihilation of Weyl nodes and opening of a band gap, there are several other scenarios in which $n_0$ can change in such systems.
Firstly, Hu \textit{et al.}~\cite{Hu_2017_2} found that the apparent phase of this oscillation is magnetic field-dependent below $\sim$4 T, and smoothly changes from a value of 0.3 below 4 T to a value of 0.6 above 4 T.
For the present analysis, all resolvable oscillations occur between 4-9 T.
It is possible that, when pressure is applied, the field at which this value of $n_0$ saturates is larger than 4 T, but it seems unlikely this would cause $n_0$ to abruptly change by a value of 0.5.
Recently, Wang \textit{et al.}~\cite{Wang_2016_3} showed that the phase factor in Weyl semimetals can be strongly dependent on the position of the chemical potential when the chemical potential is in the vicinity of the Lifshitz point.
They find that moving the chemical potential through the Lifshitz point could produce a change in the phase from 5/8 to 1/8, which could both be considered topologically nontrivial phases.
This should produce a nonmonotonic change in the phase as the chemical potential is moved past the Lifshitz point, as well as a monotonic change in the oscillation frequency.
Our data clearly show a monotonic change in the phase as well as a nonmonotonic change in the oscillation frequency.
Thus, our data is not consistent with the nontrivial-nontrivial transition described in~\cite{Wang_2016_3}. Lastly, we observe no evidence of direct Zeeman splitting in our low frequency oscillations, which would complicate the determination of the phase~\cite{Hu_2017_2}.

As for the higher frequency set of oscillations, various reports disagree on whether to consider this orbit to be topologically trivial or nontrivial - though they  agree that the SdH phase of this orbit is zero.
References \cite{Singha_2017,Matusiak_2017,Hu_2017_2,Ali_2016} consider this to be nontrivial, while references \cite{Wang_2016_1,Zhang_2017} consider it to be trivial.
Recently, Li \textit{et al.} have showed that this Fermi pocket encloses a nodal line, proving that this orbit is nontrivial with a Berry phase of $\pi$~\cite{Li2018}.

Multiple theoretical efforts concerning pressure-induced topological phase transitions have described what one can expect to observe during such a transition.
The compound LaSb has been predicted to undergo a transition from topologically trivial to nontrivial near 3-4 GPa without breaking any symmetry, which could be verified in transport experiments by observing a change in the Berry phase from 0 to $\pi$~\cite{Guo_2017_1}.
It has been predicted that a topological transition from normal insulator to topological insulator might occur in noncentrosymmetric BiTeI under moderate pressures~\cite{Bahramy_2012_1}.
Evidence of this transition has been observed in the quantum oscillation phase of one of the bulk Fermi surface oscillations~\cite{Park_2015_1}.
Liu \textit{et al.}~\cite{Liu_2014_3} showed that a Weyl semimetal phase might exist in BiTeI for a non-zero range of pressures, but this has not yet been experimentally verified, most likely due to the small range of pressure over which this phase exists. 

ZrSiSe and ZrSiTe have also been shown to possess nodal Fermi arcs, which have been observed the bulk of ZrSiS \cite{Hu2016_1,Fu2017}.
The topological phases in the ZrSiX family of materials show a transition from nodal-line to nodeless gapped phase by tuning the chalcogenide from S to Te~\cite{Hosen2017_1}. A study of these compounds under pressure might yield further insights into the nature of the transition.
Finally, an investigation of ZrSiO, which has been predicted to be a 3D weak topological insulator at ambient pressure, would be a natural next step in the investigation of this family of materials \cite{Xu2015,Onken1964}. 

\section{Conclusions}
High-pressure electrical transport measurements were performed on single crystals of the topological nodal line semimetal ZrSiS.
Measurements of SdH oscillations up to $\sim$2.2 GPa and 9 T show two oscillation frequencies.
The effective mass of the larger Fermi surface decreases, and the phase remains topologically nontrivial and roughly constant as a function of pressure.
For the smaller orbit, we find a clear change in the phase of the quantum oscillations between 0.16-0.5 GPa, which is accompanied by an abrupt decrease in the oscillation frequency.
These changes are consistent with a pressure-driven topological quantum phase transition in which a bulk band gap is introduced~\cite{Hosen2017_1,Fang_2015}.
Higher pressure measurements to \SI{20}{GPa} show no evidence for pressure-induced superconductivity down to \SI{1.8}{K}.

The apparent topological transition in ZrSiS occurs under modest pressures below \SI{0.5}{GPa}.
This very low pressure makes it possible to study the transition using a wide variety of probes that are unavailable at higher pressures.
It would be particularly interesting to see if computational efforts can shed further light on the nature of the transition.

\section*{Acknowledgments}
This work was supported by National Science Foundation (NSF) CAREER award DMR-1453752.  High pressure technique development was partially supported by a National High Magnetic Field Laboratory User Collaboration Grant.  The National High Magnetic Field Laboratory is supported by the NSF via Cooperative agreement No.\ DMR-1157490, the State of Florida, and the U.S. Department of Energy. Designer diamond anvils were supported by DOE-NNSA Grant No.\ DE-NA0002928 and under the auspices of the U.S. Department of Energy by Lawrence Livermore National Laboratory under Contract DE-AC52-07NA27344.  We thank Yuxuan Wang (UFL) for informative conversations.

\bibliographystyle{apsrev4-1}
\bibliography{ZrSiS}

\end{document}